\documentclass[preprint,12pt]{elsarticle}

\usepackage{textcomp,booktabs}
\usepackage[usenames,dvipsnames]{color}
\usepackage{colortbl}
\usepackage{rotating}
\definecolor{mygray}{gray}{.9}
\definecolor{mypink}{rgb}{.99,.91,.95}
\definecolor{mycyan}{cmyk}{.3,0,0,0}
\biboptions{numbers,sort&compress}
\usepackage{amssymb}
\usepackage{graphicx}
\usepackage{booktabs}
\usepackage{tabularx}
\usepackage{array}
\usepackage{lscape}
\usepackage{longtable}
\usepackage{multirow}
\usepackage{amsmath}
\usepackage{colortbl}
\usepackage{epsfig}
\usepackage{epsf}
\usepackage{amsthm}
\usepackage[table,xcdraw]{xcolor}
\newtheorem{definition}{Definition}[section]
\newtheorem{example}{Example}[section]

\usepackage{lineno}
\usepackage{subfig}

\newcommand{\PreserveBackslash}[1]{\let\temp=\\#1\let\\=\temp}
\newcolumntype{C}[1]{>{\PreserveBackslash\centering}p{#1}}
\newcolumntype{R}[1]{>{\PreserveBackslash\raggedleft}p{#1}}
\newcolumntype{L}[1]{>{\PreserveBackslash\raggedright}p{#1}}
\journal{Chaos, Solitons \& Fractals} 
\begin{document}

\begin{frontmatter}


\title{Identifying influential nodes based on fuzzy local dimension in complex networks}


\author[address1]{Tao Wen}
\author[address1]{Wen Jiang \corref{label1}}

\address[address1]{School of Electronics and Information,Northwestern Polytechnical University, Xi'an, shaanxi 710072, China}
\cortext[label1]{Corresponding author at: Wen Jiang, School of Electronics and Information, Northwestern Polytechnical University, Xi'an, Shannxi 710072, China, Tel: +86 02988431267. E-mail: jiangwen@nwpu.edu.cn; jiangwenpaper@hotmail.com.}
\begin{abstract}
How to identify influential nodes in complex networks is an important aspect in the study of complex network. In this paper, a novel fuzzy local dimension (FLD) is proposed to rank the influential nodes in complex networks, where a node with high fuzzy local dimension has high influential ability. This proposed method focuses on the influence of the distance from the center node on the local dimension of center node by fuzzy set, resulting in a change in influential ability. In order to show this proposed method's effectiveness and accuracy, four real-world networks are applied in this paper. Meanwhile, Susceptible-Infected (SI) is used to simulate the spreading process by FLD and other centrality measures, and Kendall's tau coefficient is used to describe the correlation between the influential nodes obtained by centrality and the results measured by SI model. Observing from the ranking lists and simulated results, this method is effective and accurate to rank the influential nodes.
\end{abstract}


\end{frontmatter}

\section{Introduction}

In recent researches, a great deal of systems in the real world exist in the form of networks, such as the relationship network in the social system\cite{Deville2016Scaling,Wang2015Impact}, protein network\cite{Zhao2016Protein,Hahn2016A}, scientists collaboration network\cite{Clough2014What}. Because the complex network\cite{Newman2003Newman} can show the structural complexity and the original properties of the original system, it is developing rapidly and wildly used in many natural sciences over the years. The study of complex network can help to quantify the fuzzy world, predict the development of the system\cite{Gao2016Universal,2016HongStructural}, and study the prisoner's dilemma experiments\cite{li2017punishment,WangSAe1601444,yu2016system}. Complex networks have produced a large number of practical models and these have
achieved a lot of practical results\cite{Zhang2016Exacerbated,Lu2015An,Rosenberg2017Minimal}.

The relationship between two nodes like diseases spreading\cite{Pei2013Spreading,wang2016statistical} and information dissemination\cite{Jankowski2017Balancing} in real-world networks has important significance\cite{Lu2015Evaluating}. If a node has greater influence in complex network, this node undertakes the heavier task in the exchange of information (the wider the scope of information diffusion from this node within the unit time), the greater the impact of removing this node in the network. Based on this, identifying influential nodes in complex network has been focus of the research of complex networks, which can control the spread of disease\cite{Yang2007Epidemic} and identify the most influential spearders\cite{Morone2017Model}. A great deal of centrality measures of identifying the influential nodes have been proposed in the past years\cite{wang2017new,Lu2016Vital}. The first person to identify the influential nodes is Shimbel\cite{Shimbel1953} in 1953, and the method is that the node's centrality is the number of shortest paths which go through the selected node. Then, many methods with high accuracy are proposed to identify the influential nodes like Betweeenness centrality (BC)\cite{Freeman1979Centrality,Newman2005A}, Degree centrality (DC)\cite{Freeman1979Centrality}, Closeness centrality (CC)\cite{Freeman1979Centrality}, Eigenvector centrality (EC)\cite{Bonacich2001Eigenvector}, Local dimension (LD)\cite{Pu2014Identifying} and so on. Degree centrality (DC) is a simple method which doesn't pay attention to the global structure, so it's lack of accuracy. Betweenness centrality (BC) and Closeness centrality (CC) focus on the global structure and are more accurate comparing DC, but the computational complexity limits its application to large scale real-world complex networks. Meanwhile, CC can't be applied in the networks with disconnected nodes which means there is no shortest path between two nodes. Eigenvector centrality (EC) has greater limitations because it can't be applied to asymmetric networks, and many positions can't be chosen. Local dimension (LD) focus on the nodes whose shortest distance from the center node are less than the box-size equally. But in general, the node contribution is different from each other because of the different distance from the center node. So this method isn't accurate. Many measures\cite{Tian2017A} like semi-local centrality (SLC)\cite{Chen2012Identifying}, LeaderRank (LR) \cite{Li2014Identifying,Lue2011Leaders} and the role of neighborhood\cite{Liu2016Identify} have been proposed and the results show the efficient performance of these methods.

Fuzzy theory is proposed by L. A. Zadeh\cite{ZADEH1965338} in 1965, and fuzzy set can be more accurate description of information\cite{Pedrycz2016Designing}. Because of it's effective to describe the uncertain information, fuzzy sets has been applied in many subjects\cite{Pedrycz2015From} like Ant Colony Optimization\cite{Castillo2015New,Castillo2015A}, volatility forecasting\cite{zhengrong2017ADAC,Maciel2017Evolving}, decision making\cite{jiang2017Intuitionistic,XDWJIJIS21929}, supplier selection \cite{liuDEMATEL2017}, similarity measure\cite{feiliguo2017new}, information granules\cite{Pedrycz2016Design,Pedrycz2016From}, designing type-2 fuzzy systems\cite{Castillo2012Optimization,Castillo2011Design}, uncertainty measure\cite{Wang2017IJCCC,AnImprovedAPIN2017} and combination of D-S theory\cite{zhengxianglin2017,zheng2017fuzzy}. The results show the effectiveness and accuracy of this method. The fuzzy sets has already been applied to calculate the fractal dimension by Pedryca\cite{Pedrycz2003Fuzzy}, and Castillo \emph{et al.} apply the fuzzy fractal dimension in many fields, such as time series prediction
\cite{Castillo2002Hybrid}, measuring the complexity of the sound signal\cite{Melin2007An}. Recently, some extensions of fuzzy set are presented to model more complicated situations like D numbers\cite{zhou2017dependence,dengentropy,mo2016new}. Due to the uncertainty in complex networks, fuzzy fractal dimension of complex network is proposed by Zhang \emph{et al.}\cite{Zhang2014Fuzzy}, and it has been applied in analysing images\cite{Ivanovici2011Fractal}, measuring financial data\cite{Bohdalova2010Fractal} and dealing with big data\cite{Wang2017An}. It can describe the covering ability of box more accurate and have less time consuming. Fractal dimension can depict the self-similarity and fractal properties of complex networks\cite{Song2007How}, so it has been applied in many fields\cite{zhang2016modeling,Lu2016The,zhangqi2017}. The local dimension has different values with different center nodes and different radius, and it's a great progress in dimension of complex networks. Because of the difference of the node's local dimension, Pu \emph{et al.}\cite{Pu2014Identifying} applied it to identify the influential nodes, and the ranking list of influential nodes is similar to the previous centrality measures. In this method, the nodes in the box are considered equally to the local dimension which is not precise.

In this paper, a novel centrality measure of identifying influential nodes is proposed which is based on fuzzy local dimension (FLD). The fuzzy sets can detailed describe the relationship between two nodes based on the topological distance between this two nodes, which can describe the topological structure more accurate. The dimension of complex networks can reveal the topological structure and dynamics structure which both have important influence to rank the influential nodes. Based on these, fuzzy local dimension is proposed to identify the influential nodes, and this method is applicable to most complex networks and takes less time than some measures like BC. The larger the fuzzy local dimension of a node, the greater the impact of this node, the wider the scope of information diffusion from this node within the unit time. In order to evaluate the validity and reliability of this method, four real-world complex networks have been applied in this paper to compare this method with other centrality measures. Susceptible-Infected (SI)\cite{Zhou2006Behaviors} model is used to simulate the spreading progress by different centrality measures, and Kendall’s tau coefficient is used to measure the correlation between the influential nodes ranking list obtained by centrality measures including fuzzy local dimension and the result measured by SI model. The results show this proposed method performs well.

The remainder of this paper is organized as follows. Some brief overview of complex network, fuzzy sets and centrality measures are given in Section 2. In Section 3, the method which is based on the fuzzy local dimension to identify the influential nodes is proposed. Susceptible-Infected (SI) and Kendall's tau coefficient is used to evaluate the results obtained by FLD in Section 4. Some conclusions are given in Section 5.

\section{Preliminaries}

In complex networks \emph{$G(N,V)$}, there are a collection of nodes \emph{$N{\rm{ = (1,2,}} \cdots {\rm{,n)}}$} and a collection of edges \emph{$V{\rm{ = (1,2,}} \cdots {\rm{,v)}}$}.
\subsection{The shortest path between any two nodes}
The shortest path between any two nodes can influence the properties of the complex network. The adjacency matrix of complex network \emph{$G$} can be obtained by the relationship between nodes and edges, which can be represented by \emph{$X = ({x_{ij}})$}. The case when \emph{${x_{ij}} = 1$} represents there is a connection between node \emph{$i$} and node \emph{$j$}, and \emph{${x_{ij}} = 0$} is the opposite case. The shortest distance\cite{Wei2016Multifractality} between node \emph{$i$} and node \emph{$j$} is expressed as \emph{${{\rm{d}}_{ij}}$}, and the maximum value of shortest distance from center node \emph{$i$} is expressed as \emph{$d_i^{\max }$}. It should be noted that the short path finding issue plays a very important role in network optimization \cite{xu2017modified,zhang2016physarum}.

\subsection{Centrality measures}
There are several node centrality measures that already exist, like degree centrality (DC), closeness centrality (CC), betweenness centrality (BC), eigenvector centrality (EC), local dimension (LD) which are defined as follows,

The DC\cite{Newman2003Newman} is the degree centrality of node \emph{$i$} which is expressed as \emph{${C_D}(i)$}. It equals to the number of edges connected with thia center node.

Because the degree centrality only considers the direct influence of node, without considering the influence of network structure to the information disseminating, the closeness centrality (CC) is proposed by Freeman which is expressed as \emph{${C_C}(i)$}. In general, information would disseminate in the shortest distance. If a node with the ability to forward the most of the information is lost, the network transmission efficiency would have a large impact. The more shortest distance going through node \emph{$i$}, the more importance node \emph{$i$}.
\begin{definition}
((CC)\cite{Freeman1979Centrality}). The closeness centrality \emph{${C_C}(i)$} is defined as follows,
 \begin{equation}\label{equ_CC}
{C_C}(i) = \frac{1}{{\sum\limits_{j = 1}^N {{d_{ij}}} }}
\end{equation}
where node \emph{$i$} is the center node, \emph{$j$} is the other nodes in the whole network, \emph{$N$} is the total number of nodes in this network, \emph{${{d_{ij}}}$} is shortest distance from node \emph{$i$} to node \emph{$j$}.
\end{definition}

Because the closeness centrality (CC) has high computational complexity which can't deal with the large scale network well, betweenness centrality (BC) is proposed by Freeman which is expressed as \emph{${C_B}(i)$}. The smaller the sum of the shortest distance \emph{${{\rm{d}}_{ij}}$} from center node \emph{$i$} to all other nodes in the network, the less time for information disseminating, and the more influential the node \emph{$i$}.
\begin{definition}
((BC)\cite{Newman2005A}). The betweenness centrality \emph{${C_B}(i)$} is defined as follows,
 \begin{equation}\label{equ_BC}
{C_B}(i) = \frac{{\sum\limits_{s \ne t \ne i} {{L_{st}}(i)} }}{{\sum\limits_{s \ne t} {{L_{st}}} }}
\end{equation}
where \emph{${{L_{st}}}$} is the number of shortest paths between node \emph{$s$} and node \emph{$t$} in the whole network, \emph{${{L_{st}}(i)}$} is the number of shortest paths between node \emph{$s$} and node \emph{$t$} which go through node \emph{$i$}.
\end{definition}

\begin{definition}
((EC)\cite{Bonacich2001Eigenvector}). The EC is the eigenvector centrality of node \emph{$i$} which is expressed as \emph{${C_E}(i)$}. \emph{$A$} is an $n \times n$ similarity matrix, and the eigenvector centrality \emph{${x_i}$} of node \emph{$i$} is the \emph{$i$}th entry in the normalized eigenvector which belongs to the similarity matrix \emph{$A$}.
 \begin{equation}\label{equ_EC}
Ax = \lambda x,{x_i} = u\sum\limits_{j = 1}^n {{a_{ij}}{x_j},i = 1,2, \cdots n}
\end{equation}
where \emph{$\lambda $} is the largest eigenvalue of similarity matrix \emph{$A$}, \emph{$n$} is the number of nodes, and the proportionality which is defined as follows,
\begin{equation*}\label{equ_EC_lamda}
u = \frac{1}{\lambda }
\end{equation*}
so \emph{${x_i}$} is the sum of similarity scores of the nodes which are connected to this node.
\end{definition}

The local dimension \emph{${C_{LD}}(i)$}\cite{Pu2014Identifying} pays attention to the individuals rather than only on the global information, and it's introduced detailed in the next section.

\subsection{Local dimension of complex networks}
To express the local properties of each node in the complex network, Silva \emph{et al.} have proposed local dimension. This method can observe the different scaling properties with different topological distance \emph{r} from the selected center node, and it has been proved that not only the small-world networks but also many real complex networks follows a power law distribution. It means that the radius \emph{$r$} (it can equal to \emph{$r$}) and the total number of nodes \emph{${B_i}(r)$} within the radius \emph{$r$} for each node \emph{$i$} have a power law relation as follows,
 \begin{equation}\label{equ_power law}
{B_i}(r) \sim \mu {r^{{D_i}}}
\end{equation}
where \emph{${{D_i}}$} represents the local dimension of node \emph{$i$}. It can be obtained by the slope of the double logarithmic scale fitting curves and Eq. (\ref{equ_power law}) can be rewritten as follows,
 \begin{equation}\label{equ_D_log}
{D_i} = \frac{d}{{d\log r}}\log {B_i}(r)
\end{equation}
where the radius \emph{$r$} is expanding from 1 to the maximum value of shortest distance \emph{$d_i^{\max }$} from center node \emph{$i$}, so the radius is discontinuous. Because of the discrete nature \cite{Ben2004Complex,Silva2013Local} in complex network, the derivative of Eq. (\ref{equ_D_log}) is still valid in this case and is expressed as follows,
 \begin{equation}\label{equ_D_derivative}
{D_i} = \frac{r}{{{B_i}(r)}}\frac{d}{{dr}}{B_i}(r)
\end{equation}
 \begin{equation}\label{equ_D_n/N}
{D_i} \approx r\frac{{{n_i}(r)}}{{{B_i}(r)}}
\end{equation}
where \emph{${{n_i}(r)}$} is the number of nodes whose shortest distance \emph{${{\rm{d}}_{ij}}$} from node \emph{$i$} is equal to radius \emph{$r$}, \emph{${B_i}(r)$} is the number of nodes whose shortest distance \emph{${{\rm{d}}_{ij}}$} from node \emph{$i$} is less than or equal to radius \emph{$r$}. The local dimension \emph{${C_{LD}}(i)$} of node \emph{$i$} is the slope of the double logarithmic scale fitting curves (\emph{$\log {B_i}({r_t})$} vs \emph{${\log {r_t}}$}) by liner regression. The smaller the local dimension \emph{${C_{LD}}(i)$}, the more importance the node \emph{$i$}.

\begin{example}
A simple network can explain how local dimension can be obtained. The network is shown as Fig. \ref{figure_LD_example_network}, and the max value of shortest distance \emph{$d_i^{\max }$} from the selected center node can be obtained by its topological structure. The relationship between the number of nodes \emph{${B_i}(r)$} within radius and the radius \emph{$r$} is shown in Fig. \ref{figure_LD_example_log}. The local dimension of the center node is the slope of the liner regression in the double logarithmic scale.

\begin{figure}
\centering
\includegraphics[scale=0.6]{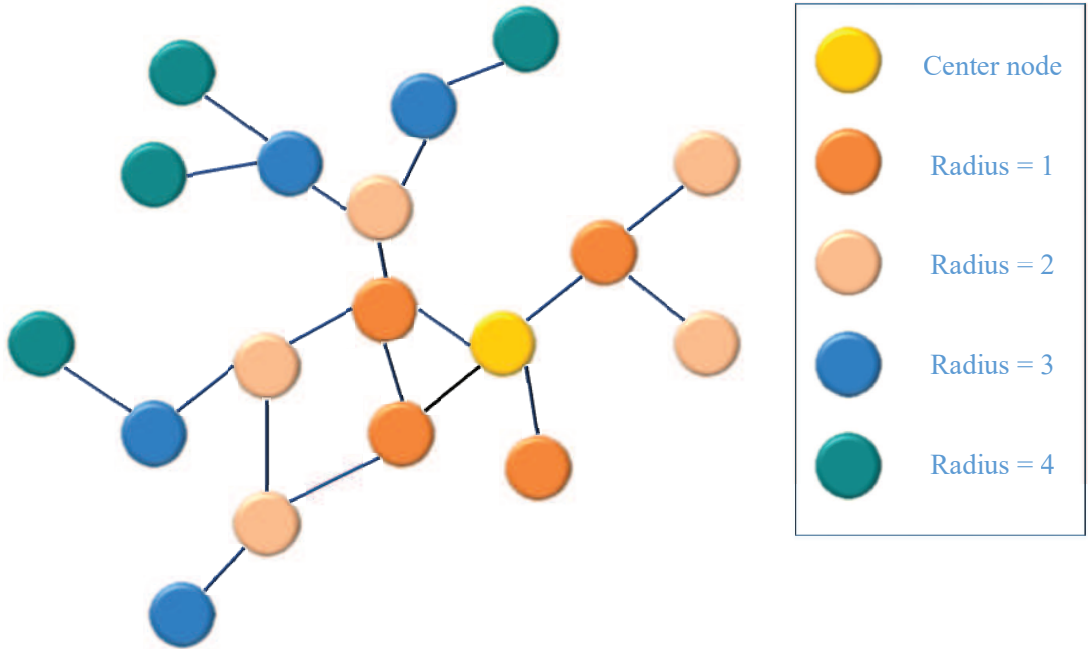}
\caption{The example network}\label{figure_LD_example_network}
\end{figure}

\begin{figure}
\centering
\includegraphics[scale=0.8]{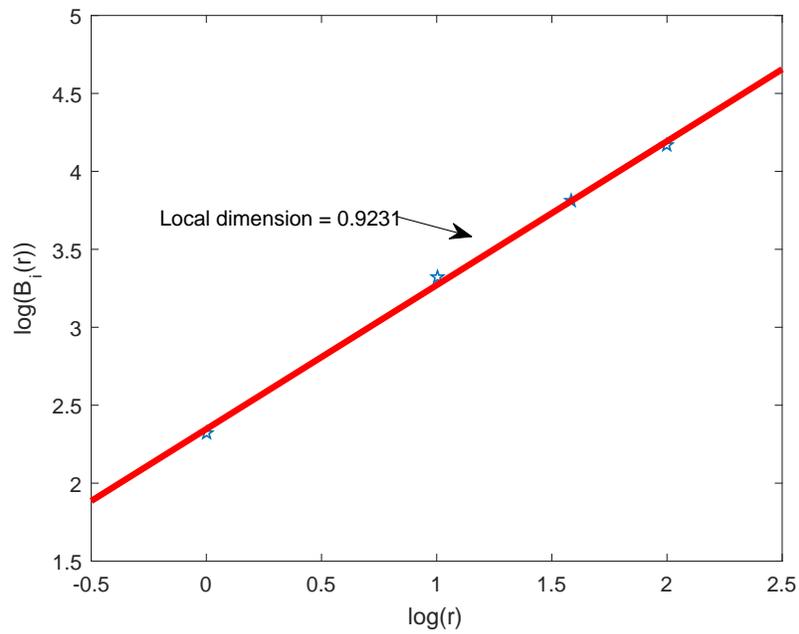}
\caption{The relationship in example network}\label{figure_LD_example_log}
\end{figure}
\end{example}

\subsection{Fuzzy sets}

In the traditional case of things divided into two categories, when there is a class \emph{$C$} which is a subset of the universal set \emph{$X$}, any case of an input variable \emph{${\rm{x}} \in X$} whether belongs to the given subset \emph{$C$} or not. There is a characteristic function \emph{${I_C}(x) \to \{ 0,1\} $} which is defined as follows,
 \begin{equation}\label{equ_mu_C}
{I_C}(x) = \left\{ {\begin{array}{*{20}{c}}
{0,}&{x \in C}\\
{1,}&{x \notin C}
\end{array}} \right.
\end{equation}
where there is a premise \emph{${\rm{x}} \in X$}.

Facing real world situations, there is no clear boundary between two categories or the boundary may be overlapping. So it's uncertain that the input variable \emph{$x$} belongs to the subset \emph{$C$} totally. To deal with this problem, this characteristic function must be improved to describe the intermediate value between 0 and 1. L. A. Zadeh\cite{ZADEH1965338} proposed the fuzzy sets to modify the characteristic function \emph{${I_C}(x)$} to the membership function \emph{${\mu _C}(x)$} which can describe the interval continuous function between 0 and 1.

\section{Fuzzy local dimension of complex networks}

\subsection{Basic method}

Based on the fuzzy sets and local dimension, a new centrality measure called fuzzy local dimension is proposed. Local dimension focus on the nodes equally whose shortest distance from the center node are less than the box-size. But in general, the node's contribution is different from each other because of the different distance from the center node. It can be easily obtained that the smaller the distance, the greater the contribution to the center node. To improve this model, fuzzy sets are used to distribute between nodes contributing to the local dimension, and each node within the box-size has a weight through the fuzzy membership function. So the fuzzy local dimension (FLD) is proposed in this paper.
\begin{definition}
(Fuzzy local dimension). Like the local dimension \emph{${D_i}$}, the fuzzy local dimension \emph{${D_{fuzzy\_i}}$} is also focus on the local properties of each node, which has the same expression as the local dimension \emph{${D_i}$} and is defined as follows,
\begin{equation}\label{equ_FLD}
{D_{fuzzy\_i}} = \frac{d}{{d\log {r_t}}}\log {N_i}({r_t},\varepsilon )
\end{equation}
because of the discrete nature \cite{Ben2004Complex,Silva2013Local} in complex network, the derivative of Eq. (\ref{equ_FLD}) is still valid in this case and is expressed as follows,
 \begin{equation}\label{equ_FLD_derivative}
{D_{fuzzy\_i}} = \frac{{{r_t}}}{{{N_i}({r_t},\varepsilon )}}\frac{d}{{d{r_t}}}{N_i}({r_t},\varepsilon )
\end{equation}
 \begin{equation}\label{equ_FLD_n/N}
{D_{fuzzy\_i}} \approx {r_t}\frac{{{n_i}({r_t})}}{{{N_i}({r_t},\varepsilon )}}
\end{equation}
where \emph{${r_t}$} is the radius from center node \emph{$i$} and it can expand from small to big in a set \emph{$\{ {\rm{1}},{\rm{2}} \cdots d_i^{\max }\} $}, \emph{${{n_i}({r_t})}$} is the fuzzy number of nodes whose shortest distance equal to the box-size \emph{$\varepsilon $}, \emph{${N_i}({r_t},\varepsilon )$} is the fuzzy number of nodes whose shortest distance is less than the box-size \emph{$\varepsilon $} which is obtained by a fuzzy set, and it can be defined as follows,
\begin{equation}\label{equ_N(LFD)}
{N_i}({r_t},\varepsilon ){\rm{ = }}\frac{{\sum\limits_{j = 1}^N {{A_{ij}}(\varepsilon )} }}{{{N_{i\_r}}}}
\end{equation}
where \emph{$\varepsilon $} is the size of box, \emph{${{N_{i\_r}}}$} is the real number of nodes when the shortest distance between node \emph{i} and \emph{j} is less than the box size \emph{$\varepsilon $}, and \emph{${A_{ij}}(\varepsilon )$} is a membership function when the distance from node \emph{$j$} to node \emph{$i$} is less than the box size \emph{$\varepsilon $} which is defined as follows,
\begin{equation}\label{equ_A}
{A_{ij}}(\varepsilon ) = \exp ( - \frac{{{d_{ij}}^2}}{{{\varepsilon ^2}}})
\end{equation}
where \emph{${d_{ij}}$} is the shortest distance between center node \emph{$i$} and node \emph{$j$}. \emph{${A_{ij}}(\varepsilon )$} follows a normal distribution, it can give the neighbour node which is relative to the center node \emph{$i$} a weight between 0 and 1 instead of a definite value of 0 or 1. It's a function of distance and it can distinguish the different nodes whose distance is different from the center node. When the nodes have different distance with center node, these nodes would have different contributions to the center node. When the neighbour node is close to the center node, it would have a bigger weight which is close to 1, and a node's weight would be a smaller value which is close to 0 when the neighbour node away from the center node. The large the membership function \emph{${A_{ij}}(\varepsilon )$} for node \emph{$i$}, the greater the contribution to the fuzzy number of nodes \emph{${N_i}({r_t},\varepsilon )$}, and node \emph{i} would be more influential.
\end{definition}

With the membership function \emph{${A_{ij}}(\varepsilon )$}, the node's contribution is not equal to the center node any more. Each node has a individual weight which is relative to the shortest distance from the center node. The closer the node is to the center node, the greater the impact on the center node, the greater the fuzzy number of nodes \emph{${N_i}({r_t},\varepsilon )$}, the more influential the center node. This proposed method is more realistic and effective. When a node is very unimportant, this node's fuzzy local dimension may be negative, and this property gives a critical basis for identifying the influential nodes.

\subsection{Example explanation}

\begin{figure}
  \centering
  \includegraphics[scale=0.8]{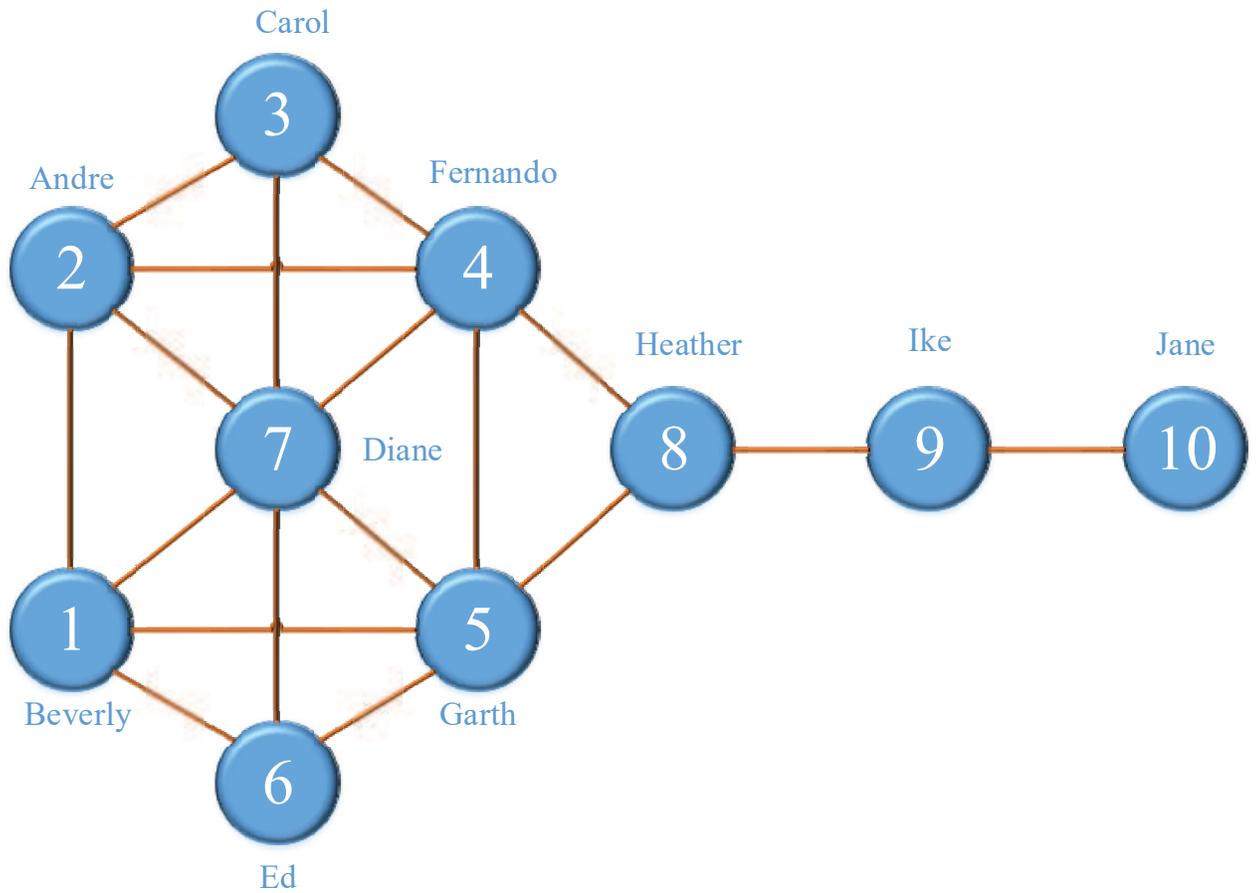}
  \caption{The Kite network}\label{figure_kite}
\end{figure}

In order to explain this proposed method more specific, a small real world complex network named as Kite network \cite{Krackhardt1990Assessing} is used to explain how this method works. This small network is a interpersonal relation network with ten nodes which is shown in Fig. \ref{figure_kite}. The Kite network describes the relationships between 10 different people, and it consists of two parts. People in left part (node 1, 2, 3, 4, 5, 6, and 7) are close to each other, and people in right part (node 8, 9, and 10) are not close.

Step 1: All the nodes in this network will be selected as the center node to calculate its fuzzy local dimension in turn. In this example, node 7 is chosen as the center node first.

Step 2 : The fuzzy number of nodes \emph{${N_i}({r_t})$} with corresponding radius \emph{${{\rm{r}}_t}$} can be obtained by Eq.(\ref{equ_N(LFD)}), where the radius \emph{${{\rm{r}}_t}$} would expand from 1 to the maximum value of shortest distance \emph{$d_7^{\max }$} from the center node 7. The fuzzy number of nodes \emph{${N_7}({r_t})$} when the radius \emph{${{\rm{r}}_t}$} equals to 1, 2, 3, and 4 can be respectively obtained as follows,

\begin{equation*}\label{equ_7_1}
\begin{array}{l}
{N_7}(1) = \frac{{\exp ( - \frac{{d_{17}^2}}{{{2^2}}}) + \exp ( - \frac{{d_{27}^2}}{{{2^2}}}) + \exp ( - \frac{{d_{37}^2}}{{{2^2}}}) + \exp ( - \frac{{d_{47}^2}}{{{2^2}}}) + \exp ( - \frac{{d_{57}^2}}{{{2^2}}}) + \exp ( - \frac{{d_{67}^2}}{{{2^2}}}) + \exp ( - \frac{{d_{77}^2}}{{{2^2}}})}}{7}\\
{\kern 1pt} {\kern 1pt} {\kern 1pt} {\kern 1pt} {\kern 1pt} {\kern 1pt} {\kern 1pt} {\kern 1pt} {\kern 1pt} {\kern 1pt} {\kern 1pt} {\kern 1pt} {\kern 1pt} {\kern 1pt} {\kern 1pt} {\kern 1pt} {\kern 1pt} {\kern 1pt} {\kern 1pt} {\kern 1pt} {\kern 1pt} {\kern 1pt} {\kern 1pt} {\kern 1pt} {\kern 1pt} {\kern 1pt} {\kern 1pt} {\kern 1pt} {\kern 1pt} {\kern 1pt}  = \frac{{\exp ( - 1) + \exp ( - 1) + \exp ( - 1) + \exp ( - 1) + \exp ( - 1) + \exp ( - 1) + \exp (0)}}{7}\\
{\kern 1pt} {\kern 1pt} {\kern 1pt} {\kern 1pt} {\kern 1pt} {\kern 1pt} {\kern 1pt} {\kern 1pt} {\kern 1pt} {\kern 1pt} {\kern 1pt} {\kern 1pt} {\kern 1pt} {\kern 1pt} {\kern 1pt} {\kern 1pt} {\kern 1pt} {\kern 1pt} {\kern 1pt} {\kern 1pt} {\kern 1pt} {\kern 1pt} {\kern 1pt} {\kern 1pt} {\kern 1pt} {\kern 1pt} {\kern 1pt} {\kern 1pt} {\kern 1pt} {\kern 1pt}  = \frac{{6 \times \exp ( - 1) + \exp (0)}}{7}\\
{\kern 1pt} {\kern 1pt} {\kern 1pt} {\kern 1pt} {\kern 1pt} {\kern 1pt} {\kern 1pt} {\kern 1pt} {\kern 1pt} {\kern 1pt} {\kern 1pt} {\kern 1pt} {\kern 1pt} {\kern 1pt} {\kern 1pt} {\kern 1pt} {\kern 1pt} {\kern 1pt} {\kern 1pt} {\kern 1pt} {\kern 1pt} {\kern 1pt} {\kern 1pt} {\kern 1pt} {\kern 1pt} {\kern 1pt} {\kern 1pt} {\kern 1pt} {\kern 1pt} {\kern 1pt}  = 0.4582
\end{array}
\end{equation*}

\begin{equation*}\label{equ_7_2}
\begin{array}{l}
{N_7}(2) = \frac{{\exp ( - \frac{{d_{17}^2}}{{{2^2}}}) + \exp ( - \frac{{d_{27}^2}}{{{2^2}}}) + \exp ( - \frac{{d_{37}^2}}{{{2^2}}}) + \exp ( - \frac{{d_{47}^2}}{{{2^2}}}) + \exp ( - \frac{{d_{57}^2}}{{{2^2}}}) + \exp ( - \frac{{d_{67}^2}}{{{2^2}}}) + \exp ( - \frac{{d_{77}^2}}{{{2^2}}}) + \exp ( - \frac{{d_{78}^2}}{{{2^2}}})}}{8}\\
{\kern 1pt} {\kern 1pt} {\kern 1pt} {\kern 1pt} {\kern 1pt} {\kern 1pt} {\kern 1pt} {\kern 1pt} {\kern 1pt} {\kern 1pt} {\kern 1pt} {\kern 1pt} {\kern 1pt} {\kern 1pt} {\kern 1pt} {\kern 1pt} {\kern 1pt} {\kern 1pt} {\kern 1pt} {\kern 1pt} {\kern 1pt} {\kern 1pt} {\kern 1pt} {\kern 1pt} {\kern 1pt} {\kern 1pt} {\kern 1pt} {\kern 1pt} {\kern 1pt} {\kern 1pt}  = \frac{{\exp ( - \frac{1}{4}) + \exp ( - \frac{1}{4}) + \exp ( - \frac{1}{4}) + \exp ( - \frac{1}{4}) + \exp ( - \frac{1}{4}) + \exp ( - \frac{1}{4}) + \exp (0) + \exp ( - 1)}}{8}\\
{\kern 1pt} {\kern 1pt} {\kern 1pt} {\kern 1pt} {\kern 1pt} {\kern 1pt} {\kern 1pt} {\kern 1pt} {\kern 1pt} {\kern 1pt} {\kern 1pt} {\kern 1pt} {\kern 1pt} {\kern 1pt} {\kern 1pt} {\kern 1pt} {\kern 1pt} {\kern 1pt} {\kern 1pt} {\kern 1pt} {\kern 1pt} {\kern 1pt} {\kern 1pt} {\kern 1pt} {\kern 1pt} {\kern 1pt} {\kern 1pt} {\kern 1pt} {\kern 1pt} {\kern 1pt}  = \frac{{6 \times \exp ( - \frac{1}{4}) + \exp (0) + \exp ( - 1)}}{8}\\
{\kern 1pt} {\kern 1pt} {\kern 1pt} {\kern 1pt} {\kern 1pt} {\kern 1pt} {\kern 1pt} {\kern 1pt} {\kern 1pt} {\kern 1pt} {\kern 1pt} {\kern 1pt} {\kern 1pt} {\kern 1pt} {\kern 1pt} {\kern 1pt} {\kern 1pt} {\kern 1pt} {\kern 1pt} {\kern 1pt} {\kern 1pt} {\kern 1pt} {\kern 1pt} {\kern 1pt} {\kern 1pt} {\kern 1pt} {\kern 1pt} {\kern 1pt} {\kern 1pt} {\kern 1pt}  = 0.7551
\end{array}
\end{equation*}

\begin{equation*}\label{equ_7_3}
\begin{array}{l}
{N_7}(3) = \frac{{6 \times \exp ( - \frac{1}{{{3^2}}}) + 1 \times \exp ( - \frac{{{2^2}}}{{{3^2}}}) + 1 \times \exp ( - \frac{{{3^2}}}{{{3^2}}}) + 1 \times \exp (0)}}{9}\\
{\kern 1pt} {\kern 1pt} {\kern 1pt} {\kern 1pt} {\kern 1pt} {\kern 1pt} {\kern 1pt} {\kern 1pt} {\kern 1pt} {\kern 1pt} {\kern 1pt} {\kern 1pt} {\kern 1pt} {\kern 1pt} {\kern 1pt} {\kern 1pt} {\kern 1pt} {\kern 1pt} {\kern 1pt} {\kern 1pt} {\kern 1pt} {\kern 1pt} {\kern 1pt} {\kern 1pt} {\kern 1pt} {\kern 1pt} {\kern 1pt} {\kern 1pt} {\kern 1pt} {\kern 1pt} {\kern 1pt}  = 0.8198
\end{array}
\end{equation*}

\begin{equation*}\label{equ_7_4}
\begin{array}{l}
{N_7}(4) = \frac{{6 \times \exp ( - \frac{1}{{{4^2}}}) + 1 \times \exp ( - \frac{{{2^2}}}{{{4^2}}}) + 1 \times \exp ( - \frac{{{3^2}}}{{{4^2}}}) + 1 \times \exp ( - \frac{{{4^2}}}{{{4^2}}}) + 1 \times \exp (0)}}{{10}}\\
{\kern 1pt} {\kern 1pt} {\kern 1pt} {\kern 1pt} {\kern 1pt} {\kern 1pt} {\kern 1pt} {\kern 1pt} {\kern 1pt} {\kern 1pt} {\kern 1pt} {\kern 1pt} {\kern 1pt} {\kern 1pt} {\kern 1pt} {\kern 1pt} {\kern 1pt} {\kern 1pt} {\kern 1pt} {\kern 1pt} {\kern 1pt} {\kern 1pt} {\kern 1pt} {\kern 1pt} {\kern 1pt} {\kern 1pt} {\kern 1pt} {\kern 1pt} {\kern 1pt} {\kern 1pt} {\kern 1pt}  = 0.8353
\end{array}
\end{equation*}

\begin{figure}
  \centering
  \includegraphics[scale=0.8]{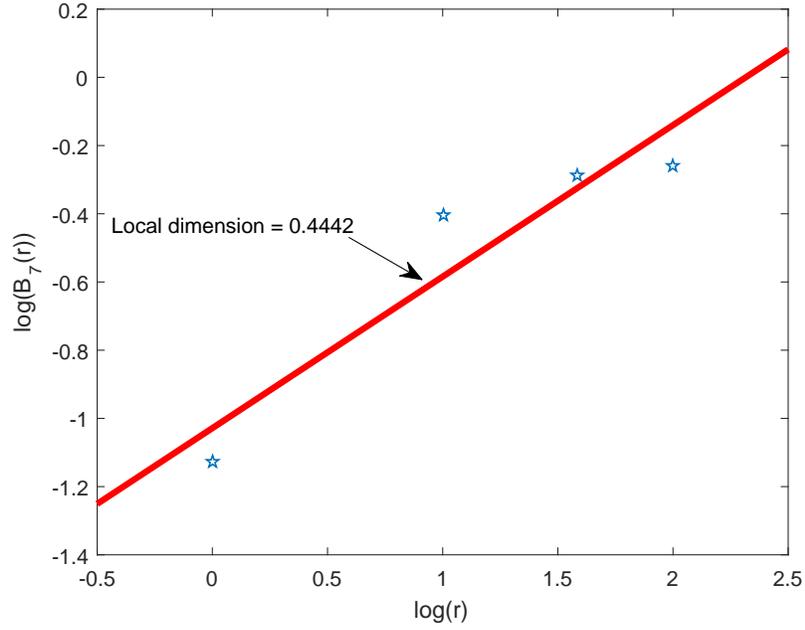}
  \caption{The double logarithmic scale fitting curves of node 7 in Kite network}\label{figure_kite_log}
\end{figure}

Step 3 : The relationship between fuzzy number of nodes \emph{${N_7}({r_t})$} and the corresponding radius \emph{${r_t}$} can be shown in Fig. \ref{figure_kite_log}. The slope of the double logarithmic scale fitting curves is the fuzzy local dimension \emph{${D_{fuzzy\_7}}$} of node 7.

Step 4 : Repeat step 1 to step 3 for all remaining nodes, and the results are shown in Table \ref{table_FLD_kite}.

\begin{table}[]
\centering
\begin{tabular}{llllll}
\hline
Node ID & 1      & 2      & 3      & 4      & 5       \\
\hline
\rowcolor[HTML]{EFEFEF}
Fuzzy local dimension  & 0.3609 & 0.3609 & 0.3015 & 0.4554 & 0.4554  \\
Node ID & 6      & 7      & 8      & 9      & 10      \\
\rowcolor[HTML]{EFEFEF}
Fuzzy local dimension  & 0.3015 & 0.4442 & 0.0760 & 0.0375 & -0.1163 \\
\hline
\end{tabular}
\caption{The fuzzy local dimension of each node in Kite network}
\label{table_FLD_kite}
\end{table}

Finally, the nodes would be sorted by the value of fuzzy local dimension, and the ranking results with this proposed method and other centrality measures are shown in Table \ref{table_Kite}. Observing from Table \ref{table_Kite} and Table \ref{table_FLD_kite}, node 4(Fernando) and node 5(Garth) play the most important roles at the same time because they are equivalent to the the door of the parts of the network, and these two nodes' fuzzy local dimension are also maximum. These two nodes connect two parts of the network. The left part of the network is more complex, so the nodes in this part are more important. Meanwhile, in the left part of Kite network, node 7(Diane) is the center of this part, so it is the most influential node in the left half of the network. The rest nodes in the left part of network including node 1(Beverly), node 2(Andre), node 3(Carol), node 6(Ed) play the same role and they are the edges of this part, the presence of these nodes makes the left part of the network more connected, so they have the same importance. Then, the right part is less important in this network, so these three nodes rank in the last three in the ranking list. But node 8(Heather) is nearest to node 4 and node 5 which are the connection nodes, node 9(Ike) is closer than node 10(Jane), so their importance ranking is 8, 9, 10. This order is the same as the ranking lists obtained by FLD, so the final ranking results from this proposed method (FLD) is more reasonable than others. A detail information in Table \ref{table_FLD_kite} is that the fuzzy local dimension of node 10 is negative, and the above conclusion shows node 10 is the least important node of Kite network. So this information shows that the fuzzy local dimension would be negative when one node is far less important than other nodes in the network.

\begin{table}[]
\centering
\begin{tabular}{ccccccc}
\rowcolor[HTML]{BBDAFF}
\hline
Rank &BC&CC&DC&EC&LD&FLD \\
\hline
1    &8 &4 &7 &7 &7 &4 \\
\rowcolor[HTML]{EFEFEF}
2    &4 &5 &5 &5 &4 &5  \\
3    &9 &7 &4 &4 &5 &7  \\
\rowcolor[HTML]{EFEFEF}
4    &5 &8 &2 &1 &1 & 2 \\
5    &7 &2 &1 &2 &2 & 1  \\
\rowcolor[HTML]{EFEFEF}
6    &2 &1 &8 &6 &3 &6 \\
7    &1 &6 &6 &3 &6 &3  \\
\rowcolor[HTML]{EFEFEF}
8    &10&3 &3 &8 &9 &8 \\
9    &6 &9 &9 &9 &10&9  \\
\rowcolor[HTML]{EFEFEF}
10   &3 &10&10&10&8 &10 \\
\hline
\end{tabular}
\caption{Kite network ranking list with different centrality measures}
\label{table_Kite}
\end{table}

\section{Experimental study}

In order to test the effectiveness and accuracy of this proposed method, four real-world complex networks have been applied for this proposed method and some existing methods such as DC, CC, BC, EC, LD. These real-world complex networks include USAir network which can be download from $(http://vlado.fmf.uni-lj.si/pub/networks/data/)$, email network which is from $(http://www.cs.bris.ac.uk/steve/peacockpaper)$, karate network which is from $(http://vlado.fmf.uni-lj.si/pub/networks/data/)$, and Groad network which is from Ref.\cite{Kaiser2004Spatial}. These networks' topological properties are shown in Table \ref{table_topological} respectively.

\begin{table}[]
\centering
\begin{tabular}{cccccccc}
\rowcolor[HTML]{BBDAFF}
\hline
Network & nodes & edges & $\left\langle k \right\rangle$ & ${k_{\max }}$ & $\left\langle d \right\rangle$  & ${d_{\max }}$ & C      \\
\hline
email   & 1133  & 10902 & 9.6222  & 71     & 3.6028  & 8      & 0.2202 \\
\rowcolor[HTML]{EFEFEF}
USAir   & 332   & 2126  & 12.8072 & 139    & 2.9299  & 6      & 0.6252 \\
karate  & 34    & 78    & 4.5882  & 17     & 2.3374  & 5      & 0.6175 \\
\rowcolor[HTML]{EFEFEF}
Groad   &  1168 & 2486  & 2.1284 &  12    &    19.4024 &  62    & 0.0012 \\
\hline
\end{tabular}
\caption{Topological properties in these four real-world complex networks. Nodes and edges represent the number of nodes and edges in these complex networks respectively; $\left\langle k \right\rangle$ and ${k_{\max }}$ show the average and maximum value of degree; $\left\langle d \right\rangle$ and ${d_{\max }}$ express the average and maximum value of shortest distance; \emph{$C$} is the clustering coefficient of these networks.}
\label{table_topological}
\end{table}

\subsection{Data}

BC, DC, CC, EC, LD and this proposed method (FLD) are used for four real-world complex networks to show the performance of this proposed method. The top-10 ranking lists of four real-world complex networks is shown in Table \ref{table_email} to Table \ref{table_Groad}. Observing from Table \ref{table_email} for email network, the proposed method has the same nine node IDs with CC; there are seven same node IDs for the proposed method (FLD) and BC, DC. In Table \ref{table_USAir} for USAir network, there are ten same node IDs for the proposed method and CC; FLD has the same nine node IDs with DC; there are six same node IDs for this proposed method and BC. In table \ref{table_karate} for karate network, the proposed method (FLD) has the same eight node IDs with BC; FLD has the same seven and six node IDs with DC and CC. In Table \ref{table_Groad} for Groad network, there are nine same node IDs for FLD and CC; there are some differences between this proposed method and BC, DC, and this proposed method has four and two same node IDs with BC and DC respectively. In Table \ref{table_karate} and Table \ref{table_Groad}, there is a difference between FLD and EC. That's because these two methods have their own emphasis on the ranking results. EC pays more attention to the whole structure of complex networks, but FLD focuses on the each individuality rather than the whole structure of complex networks. Even if their respective priorities are different, the first few nodes of the ranking result are the same, and the difference of the ranking results obtained by FLD and other centrality measures is relative small. These show that FLD can get a relatively correct influential ranking lists.

This proposed method is better than LD because there are more same nodes with other centrality measures like BC, DC, CC. In email network, FLD has two and four more same node IDs than LD with BC, CC respectively; this proposed method has one more node ID than LD with DC.In USAir network, this proposed method has four and two more same node IDs than LD with CC, BC respectively; this proposed method has equal same node ID with LD comparing with DC. In karate network, FLD has one more node ID than LD with BC, CC, DC. In Groad network, FLD has four more same node IDs than LD with CC, this proposed method has equal same node ID with LD comparing with BC, and there are one less node ID than LD with DC. From the above results, a conclusion can be easily obtained that this proposed method (FLD) is more similar than LD comparing the existed centrality measures.

\begin{table}[]
\centering
\begin{tabular}{ccccccc}
\rowcolor[HTML]{BBDAFF}
\hline
Rank & BC  & CC  & DC  & EC  & LD  & FLD \\
\hline
1    & 23  & 333 & 105 & 105 & 105 & 333 \\
\rowcolor[HTML]{EFEFEF}
2    & 105 & 23  & 333 & 16  & 333 & 23  \\
3    & 333 & 105 & 42  & 196 & 23  & 42  \\
\rowcolor[HTML]{EFEFEF}
4    & 76  & 42  & 23  & 204 & 42  & 105 \\
5    & 42  & 41  & 16  & 42  & 16  & 76  \\
\rowcolor[HTML]{EFEFEF}
6    & 578 & 76  & 41  & 49  & 434 & 468 \\
7    & 135 & 233 & 196 & 56  & 41  & 41  \\
\rowcolor[HTML]{EFEFEF}
8    & 41  & 52  & 233 & 116 & 14  & 233 \\
9    & 52  & 135 & 76  & 333 & 468 & 52  \\
\rowcolor[HTML]{EFEFEF}
10   & 355 & 378 & 21  & 3   & 299 & 378 \\
\hline
\end{tabular}
\caption{Email network}
\label{table_email}
\end{table}

\begin{table}[]
\centering
\begin{tabular}{ccccccc}
\rowcolor[HTML]{BBDAFF}
\hline
Rank & BC  & CC  & DC  & EC  & LD  & FLD \\
\hline
1    & 118 & 118 & 118 & 118 & 118 & 118 \\
\rowcolor[HTML]{EFEFEF}
2    & 8   & 261 & 261 & 261 & 261 & 261 \\
3    & 261 & 67  & 255 & 255 & 152 & 67  \\
\rowcolor[HTML]{EFEFEF}
4    & 47  & 255 & 182 & 182 & 230 & 255 \\
5    & 201 & 201 & 152 & 152 & 255 & 201 \\
\rowcolor[HTML]{EFEFEF}
6    & 67  & 182 & 230 & 230 & 182 & 182 \\
7    & 313 & 47  & 166 & 112 & 112 & 166 \\
\rowcolor[HTML]{EFEFEF}
8    & 13  & 248 & 67  & 67  & 147 & 47  \\
9    & 182 & 166 & 112 & 166 & 166 & 248 \\
\rowcolor[HTML]{EFEFEF}
10   & 152 & 112 & 201 & 147 & 293 & 112 \\
\hline
\end{tabular}
\caption{USAir network}
\label{table_USAir}
\end{table}

\begin{table}[]
\centering
\begin{tabular}{ccccccc}
\rowcolor[HTML]{BBDAFF}
\hline
Rank & BC  & CC  & DC  & EC  & LD  & FLD \\
\hline
1    & 1    & 1  & 34  & 34  & 34  & 1 \\
\rowcolor[HTML]{EFEFEF}
2    & 3    & 3  & 1   & 1   & 1   & 34 \\
3    & 34   & 34 & 33  & 3   & 33  & 33  \\
\rowcolor[HTML]{EFEFEF}
4    & 33   & 32 & 3   & 33  & 24  & 3 \\
5    & 32   & 33 & 2   & 2   & 3   & 2 \\
\rowcolor[HTML]{EFEFEF}
6    & 6    & 14 & 32  & 9   & 2   & 32 \\
7    & 2    & 9  & 4   & 14  & 30  & 24 \\
\rowcolor[HTML]{EFEFEF}
8    & 28   & 20 & 24  & 4   & 6   & 28  \\
9    & 24   & 2  & 14  & 32  & 7   & 31 \\
\rowcolor[HTML]{EFEFEF}
10   & 9    & 4  & 9   & 31  & 28  & 30 \\
\hline
\end{tabular}
\caption{karate network}
\label{table_karate}
\end{table}

\begin{table}[]
\centering
\begin{tabular}{ccccccc}
\rowcolor[HTML]{BBDAFF}
\hline
Rank & BC  & CC  & DC  & EC  & LD  & FLD \\
\hline
1    & 219 & 698 & 693 & 219 & 219 & 698 \\
\rowcolor[HTML]{EFEFEF}
2    & 543 & 219 & 403 & 217  & 369& 450  \\
3    & 758 & 450 & 300 & 267 & 217 & 565  \\
\rowcolor[HTML]{EFEFEF}
4    & 693 & 565 & 758 & 198 & 565 & 219 \\
5    & 886 & 331 & 410 & 207  & 693& 763  \\
\rowcolor[HTML]{EFEFEF}
6    & 698 & 763 & 373 & 331  & 267& 267 \\
7    & 735 & 267 & 217 & 565  & 295& 663  \\
\rowcolor[HTML]{EFEFEF}
8    & 403 & 729 & 556 & 236 & 207 & 331 \\
9    & 565 & 663 & 331 & 295 & 758 & 295  \\
\rowcolor[HTML]{EFEFEF}
10   & 763 & 347 & 219 & 231 & 331 & 729 \\
\hline
\end{tabular}
\caption{Groad network}
\label{table_Groad}
\end{table}

\subsection{Measuring effectiveness by SI model}

The standard of susceptible-infected (SI) model is used to test the different centrality measures result and capture the nodes' influential rank list. In SI model, these nodes are separate into two parts: (1)Susceptible, the number of susceptible nodes at time \emph{$t$} is represented by \emph{$S(t)$}. If one node is susceptible to this disease, this node would be affected by the neighbor nodes to become infected in the next moment. (2)Infected, the number of infected nodes at time \emph{$t$} is represented by \emph{$I(t)$}. If a node is infected, this node has the ability to affect the neighbor nodes. The initial infected nodes \emph{$I(0)$} are the test nodes which have been already gotten by previous centrality measures or random given nodes. It can be easily obtained that $S(t) + I(t) = N$, where \emph{$N$} the total number of nodes in complex networks. At each time \emph{$t$}, every infected nodes have the spreading rate \emph{$\lambda $} which is defined as $\lambda  = {(\frac{1}{2})^\beta }$ to infect their neighbor susceptible nodes, where \emph{$\beta $} has different values in different complex networks. After infection, the node in \emph{$S$} state would change to \emph{$I$} state. In general, the more influential a node, the stronger its ability to spread, the more the number of nodes infected in a certain time, the more influential this node. The number of infected nodes also can be denoted as \emph{$F(t)$}, which can be an indicator to measure the importance of the initial given infected nodes at time \emph{$t$}. It can be easily obtained that \emph{$F(t)$} would increase with the development of time \emph{$t$}, and \emph{$F(t)$} would be constant when all nodes have been infected, here it's denoted as \emph{$F({t_e})$}, where \emph{${t_e}$} is the final time when there isn't node to be infected. Obviously, the higher \emph{$F(t)$}, the more influential the initial nodes, and \emph{$F({t_e})$} represents the maximum influence of initial nodes.

\begin{figure}[]
\centering
\subfloat[]{
\begin{minipage}{8cm}
\includegraphics[width=8cm]{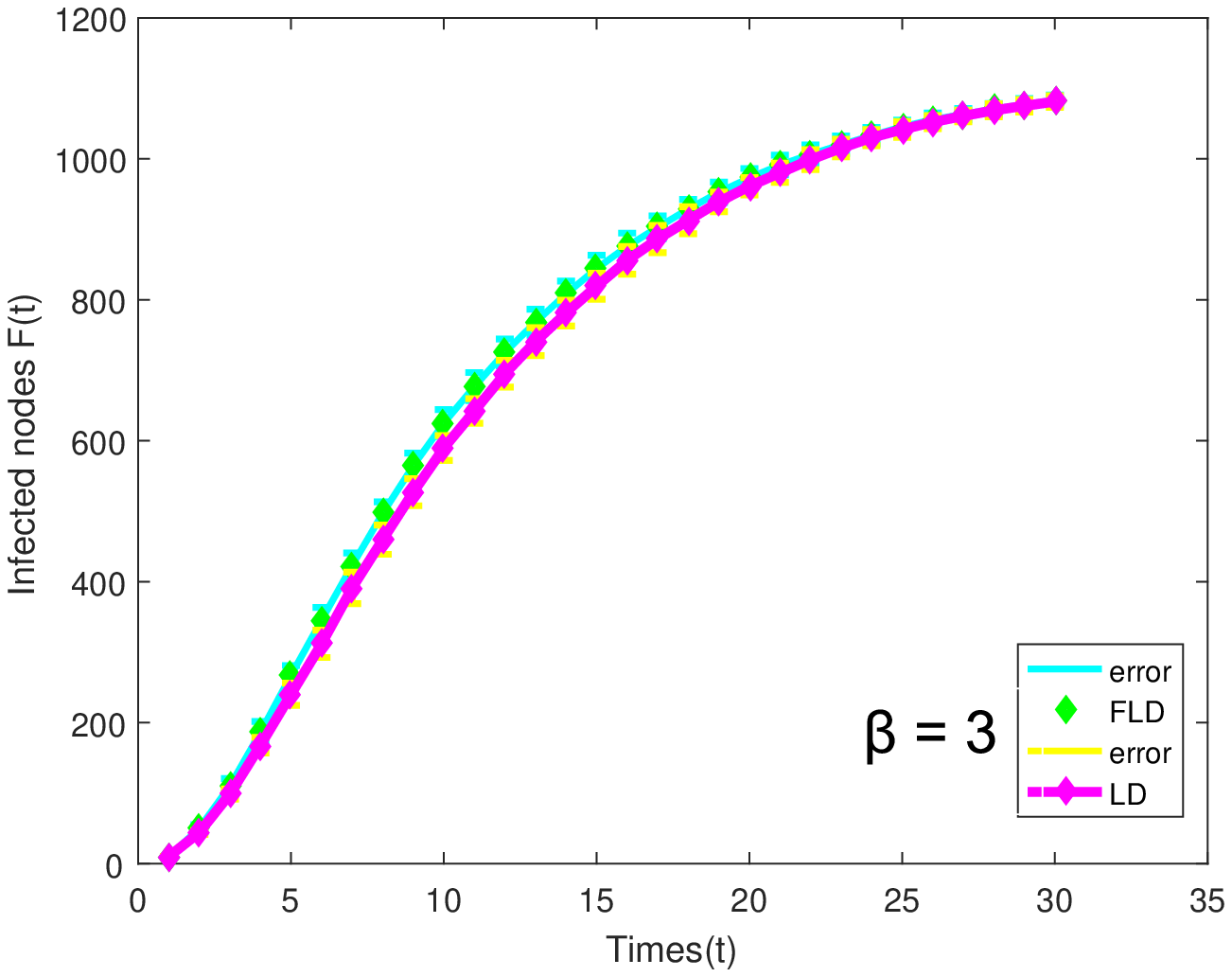}
\caption*{Email network}
\end{minipage}
}
\subfloat[]{
\begin{minipage}{8cm}
\includegraphics[width=8cm]{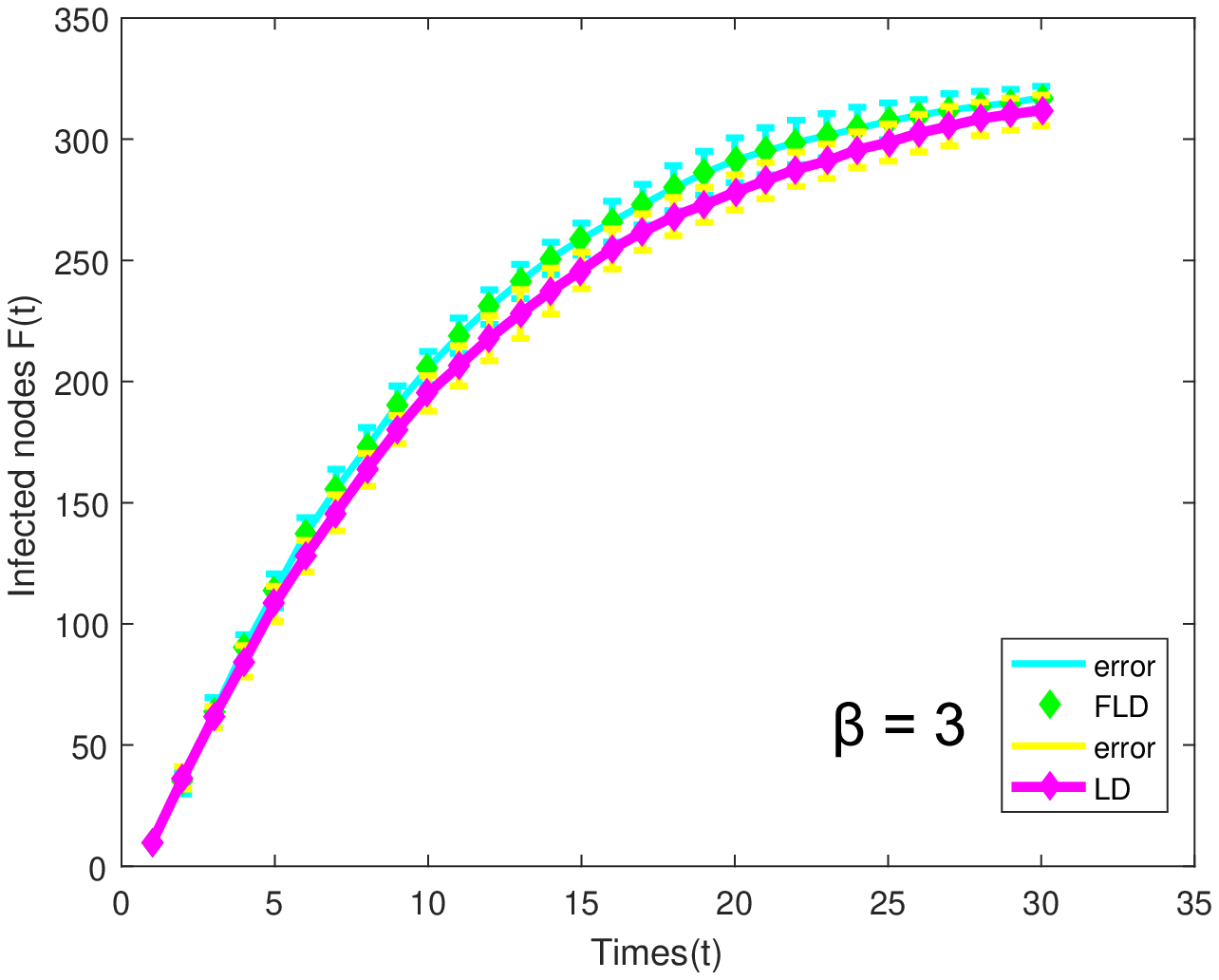}
\caption*{USAir network}
\end{minipage}
}

\subfloat[]{
\begin{minipage}{8cm}
\includegraphics[width=8cm]{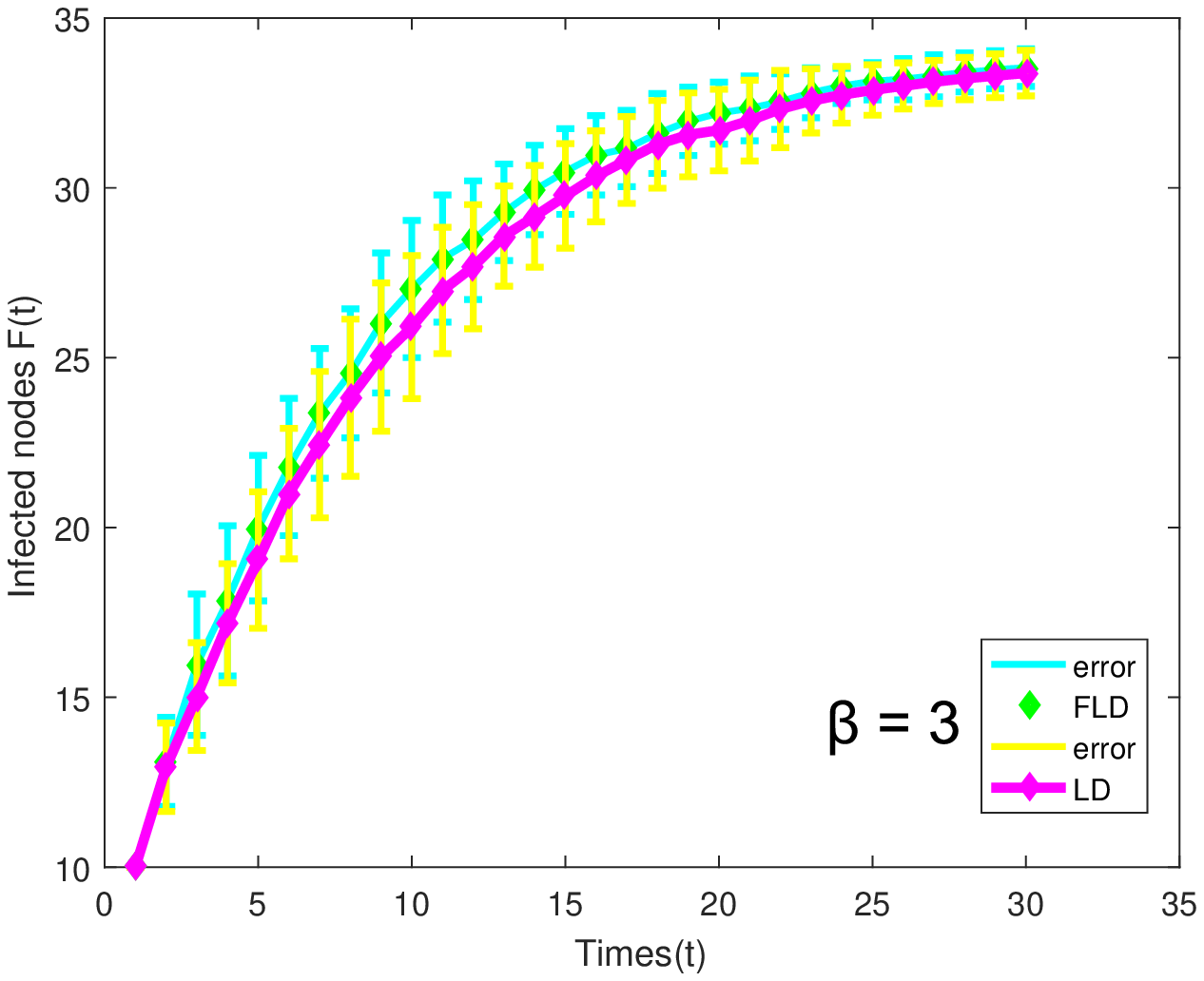}
\caption*{Karate network}
\end{minipage}
}
\subfloat[]{
\begin{minipage}{8cm}
\includegraphics[width=8cm]{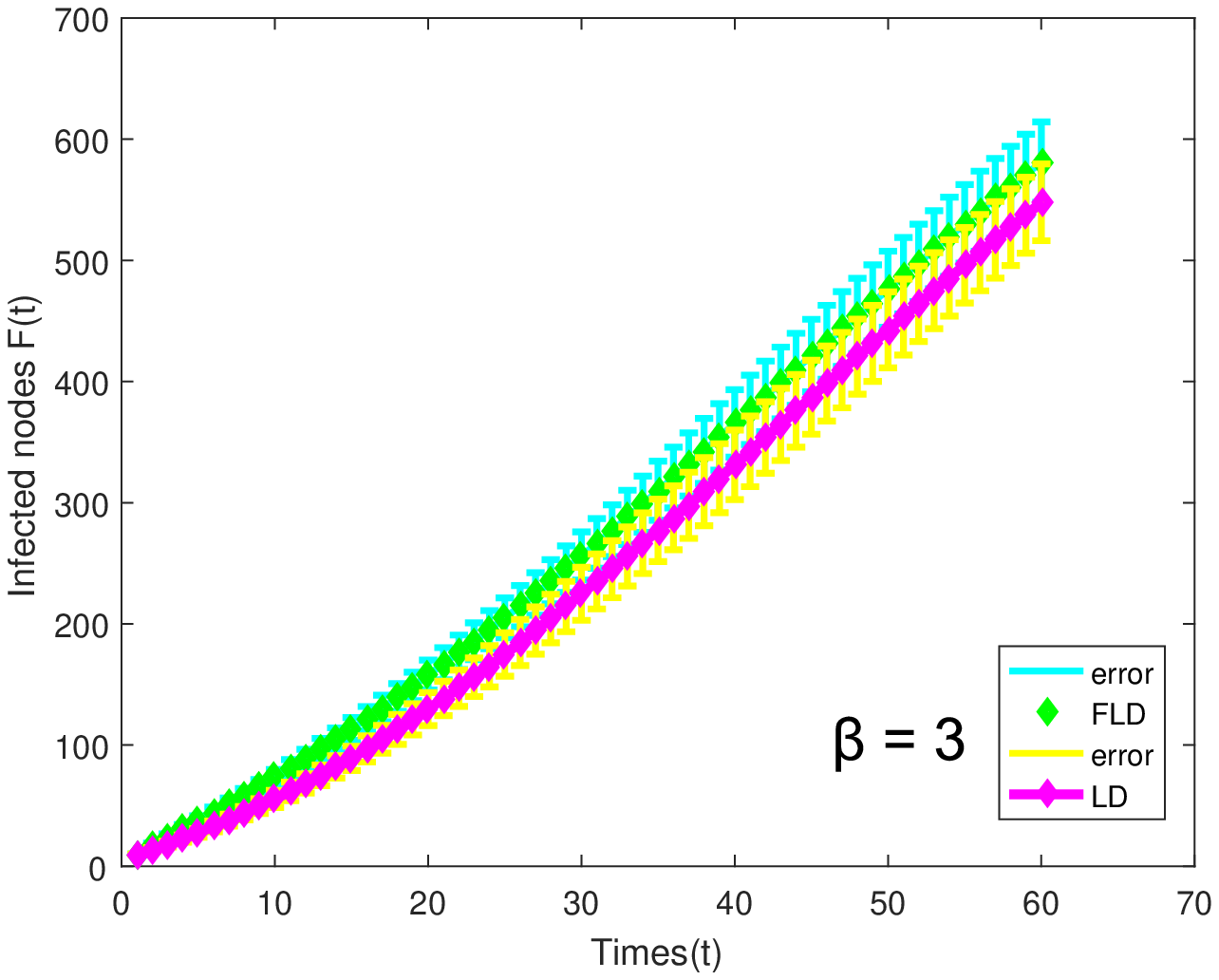}
\caption*{Groad network}
\end{minipage}
}
\caption{The number of infected nodes with different initial infected nodes (Top-10 nodes in FLD and LD) for four real-world complex networks when \emph{$\beta {\rm{ = }}3$}.}
\label{fig_SI_model}
\end{figure}

According to the property of SI model, a node which has stronger spreading ability is more important in the network. In order to compare the spreading ability of the top-10 nodes between this proposed method and local dimension which are shown in Table \ref{table_email} to Table \ref{table_Groad}, SI model is used to simulate the spreading progress of the top-10 nodes and the results of average of 100 independent tests are shown in Fig. \ref{fig_SI_model}. After the same time \emph{$t$}, if the number of infected node \emph{$F(t)$} obtained by FLD is greater than LD, it can be shown that the spreading ability of the top-10 nodes obtained by FLD is stronger than obtained by LD, and the opposite case is also true. In Fig. \ref{fig_SI_model}, the larger the number of infected node \emph{$F(t)$}, the larger the vertical value is, the higher the curve is. In order to ensure the accuracy of the experiment and to prevent contingency, the simulation process with different top-10 infected nodes would be performed 100 times. So the curve represents the mean value of 100 experiments, and the error represents the possible interval at each time \emph{$t$}.

For Email network shown in Fig. \ref{fig_SI_model}(a), the initial nodes in this proposed method have stronger spreading ability than LD, because the number of infected nodes \emph{$F(t)$} by FLD is larger than LD in the middle infecting progress. For USAir network shown in Fig .\ref{fig_SI_model}(b), the number of infected nodes \emph{$F(t)$} by this proposed method is larger than LD in whole progress, so this proposed method performs better than LD. For Karate network and Groad network shown in Fig .\ref{fig_SI_model}(c) and Fig .\ref{fig_SI_model}(d), although the error of average of 100 independent tests is larger than Email network and USAir network, the number of infected nodes \emph{$F(t)$} by FLD is larger than LD obviously. So this proposed method outperforms LD. Through these results, an obvious conclusion can be obtained that the initial nodes obtained by this proposed method (FLD) have stronger spreading ability than LD, so FLD is more useful to identify the influential nodes in complex networks.

\subsection{Measuring effectiveness by Kendall's tau coefficient}

To measure the correlation between the influential nodes obtained by centrality measures including fuzzy local dimension and the result measured by SI model, Kendall's tau coefficient\cite{Leslie1945THE} which is applied in many fields\cite{L2016Vital} is used in this paper. The higher the Kendall's tau coefficient, the more similar the two sequences, the more accurate this proposed method.

Two random variables W and V are assumed, and their \emph{$i$}th combination is represented as \emph{$({W_i},{V_i})$}. If both \emph{${W_i} > {W_j}$} and \emph{${V_i} > {V_j}$} or if both \emph{${W_i} < {W_j}$} and \emph{${V_i} < {V_j}$} occur at the same time, the \emph{$i$}th combination \emph{$({W_i},{V_i})$} and \emph{$j$}th combination \emph{$({W_j},{V_j})$} are considered concordant. If both \emph{${W_i} > {W_j}$} and \emph{${V_i} < {V_j}$} or if both \emph{${W_i} < {W_j}$} and \emph{${V_i} > {V_j}$} occur at the same time, it means that these two combinations are discordant. And if \emph{${W_i} = {W_j}$} or \emph{${V_i} = {V_j}$}, these two combinations are neither concordant nor discordant.
\begin{definition}
(Kendall's tau coefficient)Then the Kendall's tau coefficient \emph{$\tau $} is defined\cite{Hoeffding1948A,Hauskrecht2007Feature} as follows,
\begin{equation}\label{}
\tau  = \frac{{{n_c} - {n_d}}}{{0.5n(n - 1)}}
\end{equation}
where \emph{$n$} is total combinations in these sequences, \emph{${{n_c}}$} and \emph{${{n_d}}$} are the the number of concordant combinations and discordant combinations respectively. The higher the Kendall's tau coefficient \emph{$\tau $}, the more accurate the centrality measure which is used, and the smaller the \emph{$\tau $} is the opposite case. The case \emph{$\tau  = 1$} represents the ranking list by the centrality measures is same as the result obtained by SI model and the case \emph{$\tau  = 0$} is the opposite case.
\end{definition}

\begin{figure}[]
\centering
\subfloat[]{
\begin{minipage}{8cm}
\includegraphics[width=8cm]{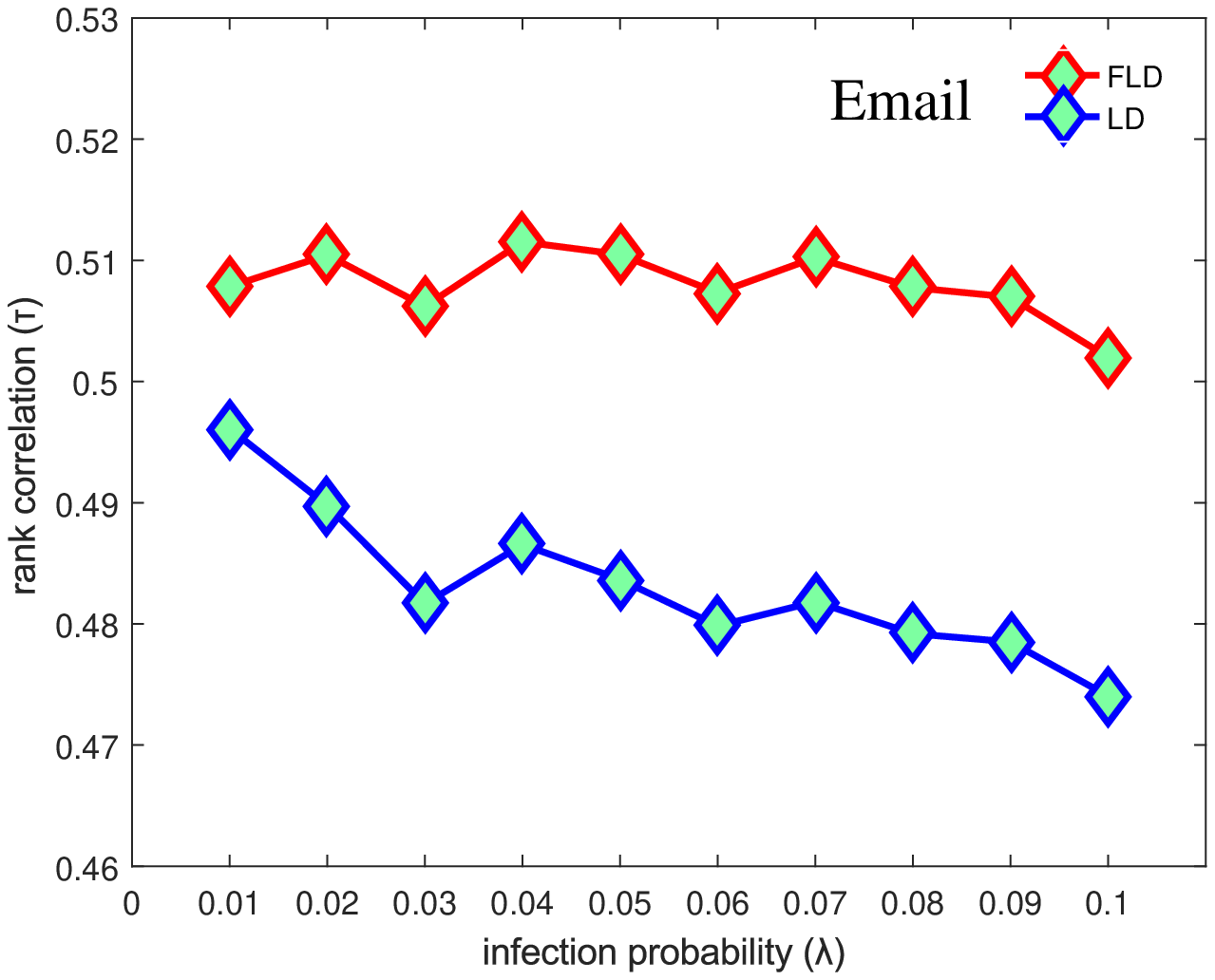}
\end{minipage}
}
\subfloat[]{
\begin{minipage}{8cm}
\includegraphics[width=8cm]{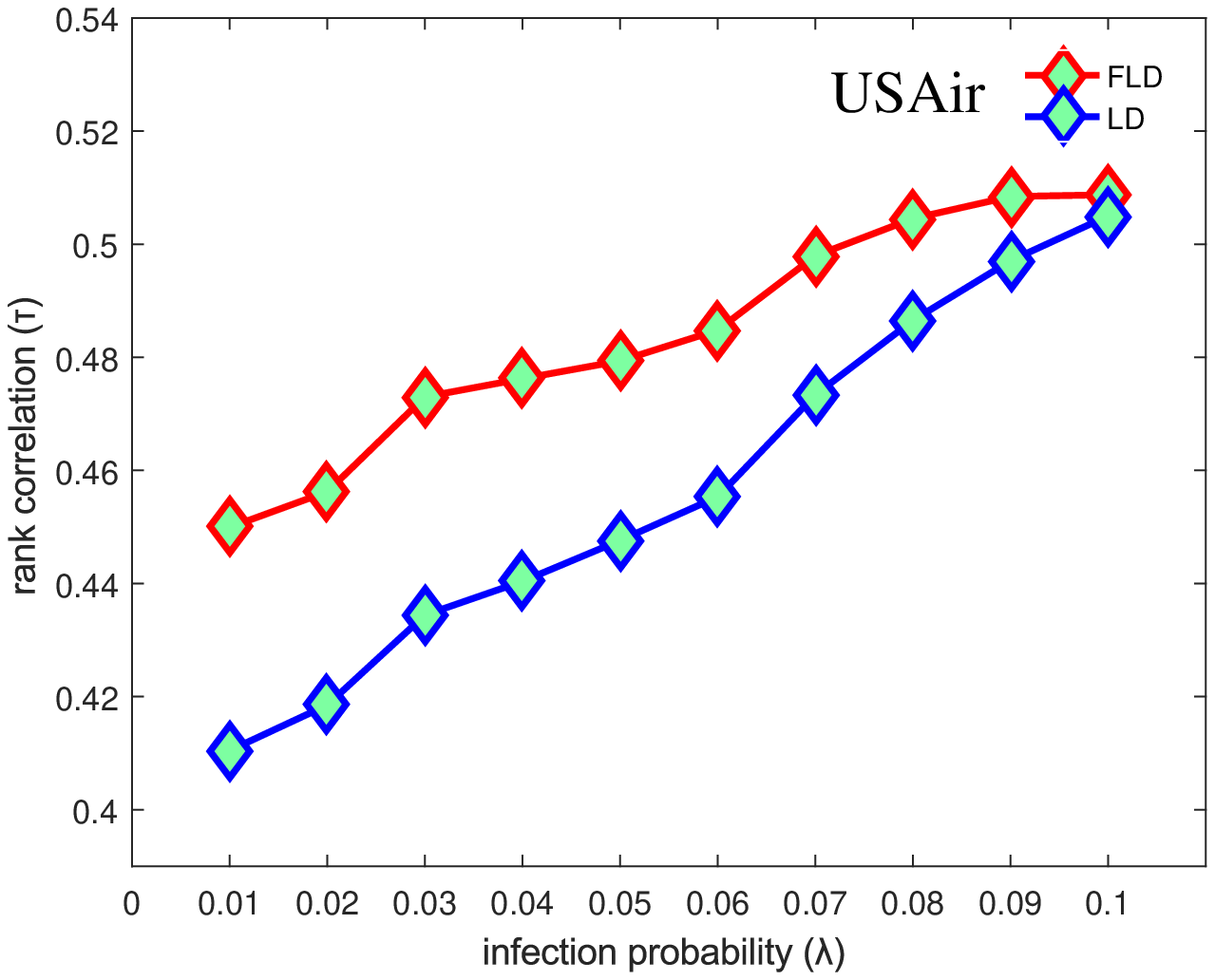}
\end{minipage}
}

\subfloat[]{
\begin{minipage}{8cm}
\includegraphics[width=8cm]{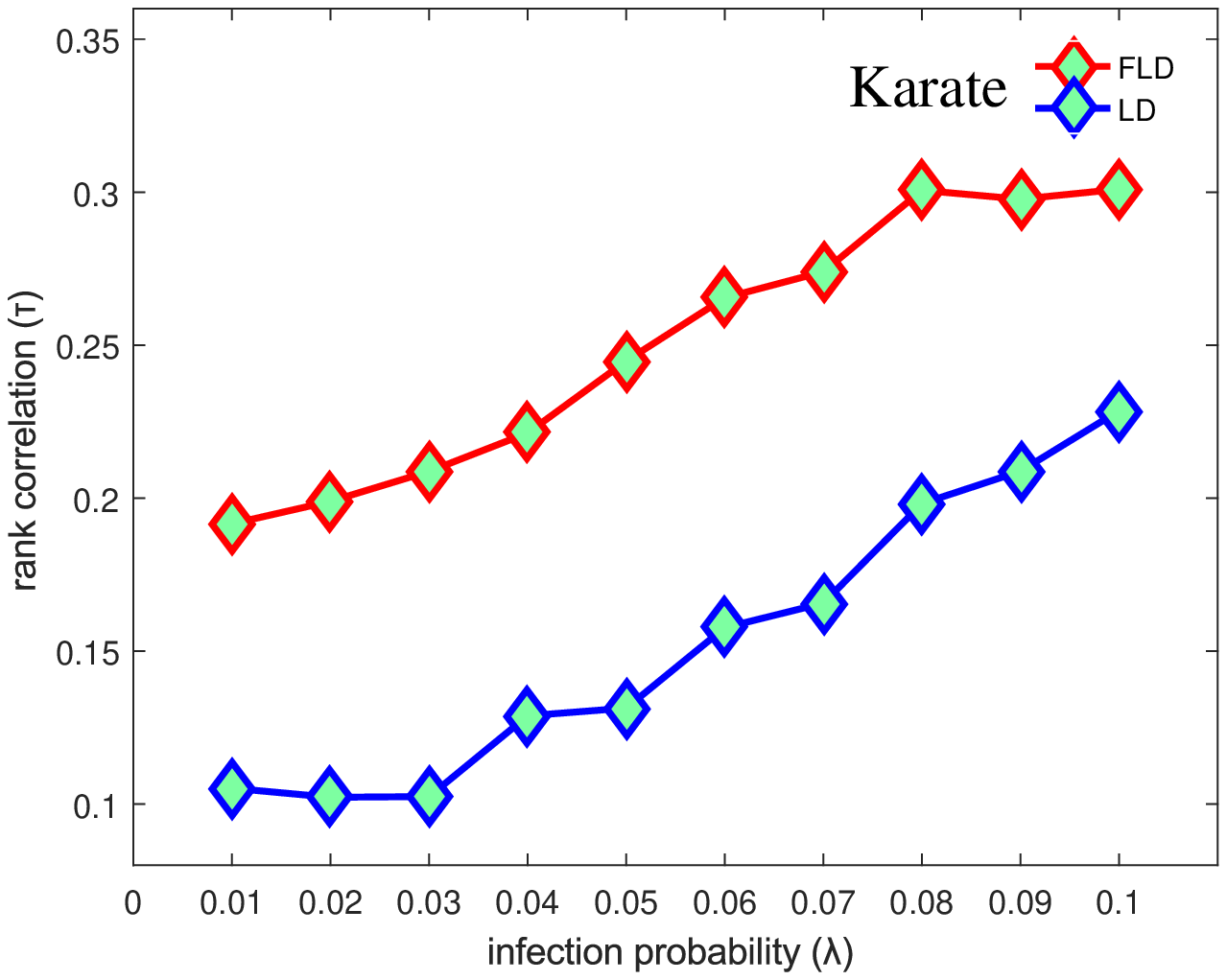}
\end{minipage}
}
\subfloat[]{
\begin{minipage}{8cm}
\includegraphics[width=8cm]{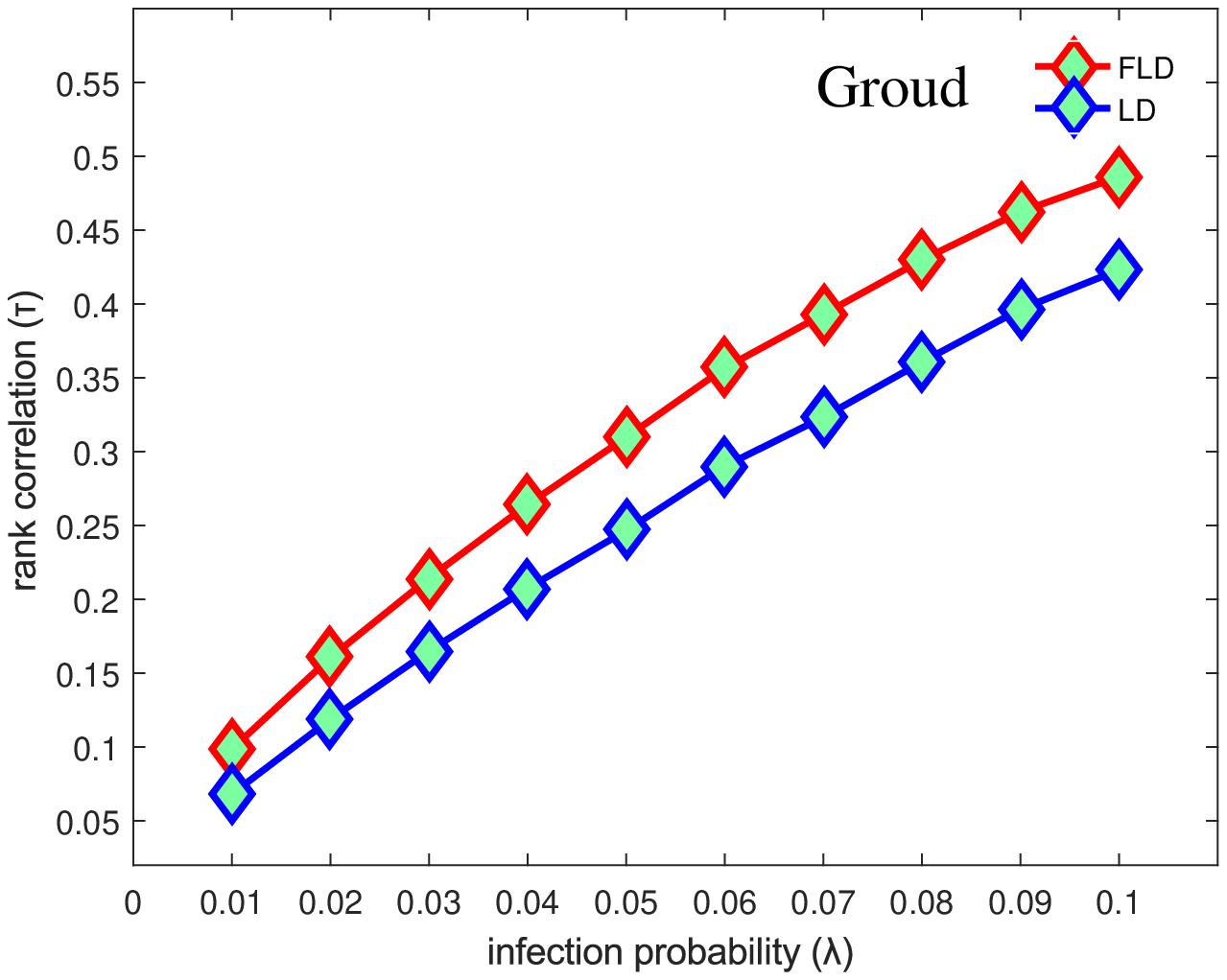}
\end{minipage}
}
\caption{The Kendall's tau coefficient \emph{$\tau $} is obtained by FLD, LD and the ranking list from SI model on different networks. The result is the average of 100 independent experiments and the spreading probability \emph{$\lambda $} is increasing from 0.01 to 0.1}
\label{fig_kendall}
\end{figure}

Kendall's tau coefficient \emph{$\tau $} can clearly show the similarity extent between two centrality measures, and it can use numerical results to intuitively show the correlation. When setting the spreading probability \emph{$\lambda $} increasing from 0.01 to 0.1, the Kendall's tau coefficient \emph{$\tau $} would increase with the increasing of spreading probability and it is shown in Fig. \ref{fig_kendall}. It's clear shown when the correlation between the centrality measures and SI model is higher, the Kendall's tau coefficient \emph{$\tau $} is larger. From Email network in Fig. \ref{fig_kendall}(a), the Kendall's tau coefficient \emph{$\tau $} does not change greatly with the increasing of spreading probability, but the Kendall's tau coefficient \emph{$\tau $} by FLD is large than LD, so this proposed method outperforms LD. In USAir network and Karate network in Fig. \ref{fig_kendall}(b) and (c), the Kendall's tau coefficient \emph{$\tau $} by FLD is much larger than LD, so this proposed method performs better than LD. For Groad network in Fig. \ref{fig_kendall}(d), the Kendall's tau coefficient \emph{$\tau $} by FLD is larger than LD and the result shows that FLD outperforms LD. The comparison between Kendall's tau coefficient \emph{$\tau $} obtained by FLD and LD shows that this FLD outperforms LD and this proposed method is more useful to identify the influential nodes than LD.

\section{Conclusion}

In this paper, a novel fuzzy local dimension is proposed to rank the influential nodes in complex networks. In this method, the nodes with different values of shortest distance from the center node have a different weight to the local dimension of center node by membership function, which is more realistic. The larger the fuzzy local dimension of this center node, the higher ranking the node in the list, the more influential this node. The positive and negative change of the fuzzy local dimension gives a clear demarcation of the ranking list of influential nodes. Compared with other centrality measures, this proposed method focus on each individuality rather than the whole structure of complex networks. Compared with LD, this proposed method takes the different distances from the center node into consideration, and each node within the box-size has a weight through the fuzzy membership function. The node which is closer to the center node would have a greater impact on the center node. Applied to four real-world complex networks, the results from the proposed method (FLD) are in good agreement with the previous centrality measures, and it performs better than LD. Susceptible-Infected (SI) model is used to simulate the spreading progress by different centrality measures, and Kendall's tau coefficient is used to measure the correlation between the influential nodes obtained by centrality measures including fuzzy local dimension and the result measured by SI model. The results of these four real-world networks applied to these two models show the effectiveness and accuracy of this method.

\section*{Acknowledgment}
The work is partially supported by National Natural Science Foundation of China (Program No. 61671384, 61703338), Natural Science Basic Research Plan in Shaanxi Province of China (Program No. 2016JM6018), Project of Science and Technology Foundation, Fundamental Research Funds for the Central Universities (Program No. 3102017OQD020).

\section*{Reference}

\bibliographystyle{ieeetr}
\bibliography{myreference}

\end{document}